\documentclass[aps,prd,preprint]{revtex4-1}
\usepackage {graphicx}
\usepackage[toc,page]{appendix}
\usepackage{amsmath}
\newcommand{\be}{\begin{equation}}
\newcommand{\ee}{\end{equation}}
\begin{document}
\title{The electric field of a charge in the vicinity of a higher dimensional black hole}
\author{David Garfinkle}
\email{garfinkl@oakland.edu}
\affiliation{Dept. of Physics, Oakland University, Rochester, MI 48309, USA}
\affiliation{Leinweber Center for Theoretical Physics, Randall Laboratory of Physics, University of Michigan, Ann Arbor, MI 48109-1120, USA}

\date{\today}

\begin{abstract}
We find the electric field of a point charge in the presence of a higher dimensional black hole.  As the charge is lowered to the horizon, all higher multipole moments go to zero, and only the Coulomb field remains.

\end{abstract}


\maketitle

\section{Introduction}
A black hole has no hair.  That is, the properties of a stationary black hole in four spacetime dimensions are entirely determined by its mass, spin, and charge.\cite{israel1,israel2,robinson,mazur}
When objects fall into a black hole, the black hole settles down to this simple, unique, stationary state.  A nice illustration of this phenomenon is contained in the paper of Cohen and Wald \cite{wald}, which calculates the electric field of a static point charge in the presence of a Schwarzschild black hole.  While this paper contains a detailed expression for the electric field, its main result is that as the position of the charge approaches the event horizon all higher multipole moments of the electric field go to zero, and only the Coulomb field remains.  

In more than four spacetime dimensions, there are many more exotic possibilities for black holes. (for a review see \cite{hollands}).  Nonetheless, for static black holes the theorems of \cite{israel1,israel2} generalize.\cite{gibbons} A static, vacuum, asymptotically flat black hole in $n+1$ spacetime dimensions is the Schwarzschild-Tangherlini black hole.\cite{tangherlini}  In the electrovac case, it is the charged generalization of the Schwarzschild-Tangherlini black hole.

Given the uniqueness result of \cite{gibbons} one would expect the result of \cite{wald} to generalize to higher dimensions.  This issue was addressed by Fox,\cite{fox} who considers the problem of a point charge in the presence of a Schwarzschild-Tangherlini black hole.  The claimed result of \cite{fox} is that in contrast to the $3+1$ dimensional case, the higher multipoles do not go away as the charge is lowered to the horizon.

In this paper, we calculate the electric field of a point charge in the presence of a Schwarzschild-Tangherlini black hole.  In contrast to \cite{fox} we find that the higher multipole moments vanish as the charge is lowered to the horizon, just as they do in \cite{wald}.  The calculation of the electric field is given in section \ref{field}, with some of the details of the calculation provided in section \ref{radial}. Conclusions are given in section \ref{conclusion}.

\section{Field calculation}
\label{field}

The line element of the Schwarzschild-Tangherlini black hole in $n+1$ spacetime dimensions takes the form
\begin{equation}
d {s^2} = - f d {t^2} \, + \, {f^{-1}} d {r^2} \, + \, {r^2} (d{\theta ^2} + {\sin^2} \theta {\gamma _{AB}} d {x^A} d{x^B} ) \; \; \; .
\label{trmetric}
\end{equation}
Here the quantity in parentheses is the line element of the $n-1$ dimensional sphere, with $\gamma _{AB}$ being the metric of the $n-2$ dimensional sphere.  The reason for writing the metric in this way is that we will choose the position of the charge to be the $z$ axis, and will thus consider functions depending only on $r$ and $\theta$.  The quantity $f$ is given by 
\begin{equation}
f = 1 \; - \; {\frac {2 M} {r^{n-2}}}
\end{equation}

For the most part, our treatment will be a straightforward generalization of the treatment in \cite{wald}, with one exception: we will begin by choosing a different set of coordinates.  The reason for this is that the $t$ coordinate is singular on the horizon.  Therefore imposing smoothness conditions on tensor fields using the coordinate system of eqn. (\ref{trmetric}) must involve careful calculation of the behavior of invariant quantities.  In contrast, given a smooth coordinate system, all that is needed is to check that the coordinate components of the relevant tensor fields are smooth functions of the coordinates.  We will choose ingoing Eddington coordinates \cite{eddington} (sometimes called Eddington-Finkelstein coordinates \cite{finkelstein}) given by 
\begin{equation}
dv = dt + {f^{-1}} dr
\end{equation}
This puts the line element of eqn. (\ref{trmetric}) in the form
\begin{equation}
d {s^2} =  - f d {v^2} \, + \, 2 dv dr   \, + \, {r^2} (d{\theta ^2} + {\sin^2} \theta {\gamma _{AB}} d {x^A} d{x^B} ) \; \; \; .
\label{vrmetric}   
\end{equation} 
In terms of metric components we have
\begin{equation}
{g_{vv}} = - f , \; \; {g_{vr}}={g_{rv}}=1 , \; \;  {g_{rr}} = 0, \; \; {g_{\theta \theta}} = {r^2}, \; \; {\sqrt g} = {r^{n-1}} {{(\sin \theta )}^{n-2}} {\sqrt \gamma}
\label{vrmetric2}
\end{equation}
The static Killing vector, $\xi ^a$ has component ${\xi ^v}=1$ with all other components vanishing.

For the electrostatic field of a point charge on the $z$ axis, the only nonzero components of the electromagnetic field tensor $F^{ab}$ are ${F^{vr}}= - {F^{rv}}$ and ${F^{v \theta}} = - {F^{\theta v}}$ where these components are functions of only $r$ and $\theta$.  From the Maxwell equation
$ {\nabla _{[a}}{F_{bc]}} = 0 $ we obtain
\begin{equation}
0= {\partial _v}{F_{r\theta}} + {\partial _r}{F_{\theta v}} + {\partial _\theta}{F_{vr}}
\end{equation}
which using eqn. (\ref{vrmetric2}) becomes
\begin{equation}
0 = {\partial _r}( - f {r^2} {F^{\theta v}} ) + {\partial _\theta}{F^{rv}}
\end{equation}
Therefore there is a scalar $\psi$ for which
\begin{equation}
{F^{rv}} = {\partial_r}\psi , \; \; \; {F^{\theta v}} = {f^{-1}}{r^{-2}} {\partial _\theta} \psi
\label{potential}
\end{equation}
From the second Maxwell equation 
\begin{equation}
- 4 \pi {j^\beta} = {\frac 1 {\sqrt g}} {\partial _\alpha} \left ( {\sqrt g} {F^{\alpha \beta}} \right )
\end{equation}
and eqn. (\ref{potential}) we find
\begin{equation}
{\partial _r}{\partial _r}\psi + {\frac {n-1} r} {\partial _r} \psi + {r^{-2}}{f^{-1}} \left ( 
{\partial _\theta}{\partial _\theta}\psi + (n-2)\cot \theta {\partial _\theta}\psi \right ) = - 4 \pi {j^v}
\label{psi}
\end{equation}

We consider a point charge $e$ located on the $z$ axis at $r=b$.
Away from the charge, we look for solutions of eqn. (\ref{psi}) by the method of separation of variables.  That is, we seek a solution of the form $\psi = A(r) B (u)$ where $u=\cos \theta$.  We then find that eqn. (\ref{psi}) gives
\begin{eqnarray}
{\frac {{d^2} A} {d {r^2}}} + {\frac {n-1} r} \, {\frac {dA} {dr}} - {\frac K {{r^2} f}} \, A = 0
\label{Aeqn}
\\
(1 - {u^2}) {\frac {{d^2}B} {d{u^2}}} + (1-n) u \, {\frac {dB} {du}} + K B = 0
\label{Beqn}
\end{eqnarray}
where $K$ is the separation constant of the equation.  The solutions of the second of these equations are the Gegenbauer polynomials ${C^\alpha _\ell}(u)$.  Here, $\ell $ is the order of the polynomial and $\alpha = (n-2)/2$.  The separation constant is 
$K = \ell (\ell + n - 2)$.  For $n=3$ the Gegenbauer polynomials are just the usual Legendre polynomials.  The Gegenbauer polynomials are orthogonal with weight function 
${(1-{u^2})}^{(n-3)/2}$ and satisfy the normalization
\begin{equation}
{\int _{-1} ^1} \; {{({C^\alpha _\ell})}^2} \, {{(1-{u^2})}^{(n-3)/2}} du = 
{\frac {\pi {2^{1-2\alpha}} \Gamma (\ell + 2 \alpha)} {\ell ! (\ell + \alpha) {{(\Gamma (\alpha))}^2}}}
\label{norm}
\end{equation}
We will use the symbol $Q^\alpha _\ell$ to denote the somewhat complicated looking normalization constant on the right hand side of eqn. (\ref{norm}).

With the known value of the separation constant, eqn. (\ref{Aeqn}) then becomes
\begin{equation}
{\frac {{d^2} {A_\ell}} {d {r^2}}} + {\frac {n-1} r} \, {\frac {d{A_\ell}} {dr}} - {\frac {\ell (\ell + n - 2)} {{r^2} f}} \, {A_\ell} = 0
\label{Aeqn2}
\end{equation}
For each $\ell$ we must find separate solutions of eqn. (\ref{Aeqn2}), one for $r<b$ and one for $r>b$.  The solution must be continuous at $r=b$, and we will compute the discontinuity in $dA/dr$ using eqn. (\ref{psi}).  

We will treat the $\ell =0$ case separately.  Here $B=1$ and 
\begin{equation}
{\frac {{d^2} {A_0}} {d {r^2}}} + {\frac {n-1} r} \, {\frac {d{A_0}} {dr}} = 0
\end{equation}
It then follows that ${F^{v\theta}}=0$ and 
\begin{equation}
{F^{vr}} = {c_0} {r^{1-n}}
\end{equation}
where the constant $c_0$ must be chosen seperately for $r<b$ and $r>b$.  Since the black hole has no charge, we must choose ${c_0}=0$ for the $r<b$ solution.  Since the charge as calculated from the field at large distances must equal $e$, it follows from eqn. (\ref{psi}) that for the $r>b$ solution
\begin{equation}
{c_0} = - {\frac {4 \pi e} {{\cal A}_{n-1}}}
\end{equation}
Here ${\cal A}_{n-1}$ is the area of the $n-1$ sphere and is given explicitly by
\begin{equation}
{{\cal A}_{n-1}} = {\frac {2 {\pi^{n/2}}} {\Gamma (n/2)}}
\end{equation}
Thus, we find that the $\ell=0$ part of the electromagnetic field is given by 
\begin{equation} 
{F^{vr}} = 0 \; \; {\rm for} \; {r<b}, \; \; \; {F^{vr}} = - {\frac {4 \pi e} {{\cal A}_{n-1}}} {r^{1-n}}
\; \; {\rm for} \; {r>b}.
\label{Coulomb}
\end{equation}

Now we consider the $\ell >0$ part of the electromagnetic field.  Since $f \to 1$ at large $r$ it follows that the solutions of eqn. (\ref{Aeqn2}) behave like $r^\ell$ and $r^{-(\ell + n - 2)}$ at large $r$.  For $r>b$ we must choose the solution that goes to zero at large distances.  Denote this solution ${g_\ell}(r)$ with its normalization chosen so that ${g_\ell}(r)={r^{-(\ell + n - 2)}}$ at large distances.  Since $f$ vanishes on the horizon, it follows from eqn. (\ref{potential}) that in order to have a smooth electromagnetic field, the solution of eqn. (\ref{Aeqn2}) must vanish on the horizon.  Denote by ${h_\ell}(r)$ this solution, with the normalization chosen so that 
${h_\ell}(r)={r^\ell}$ at large distances.  Then the $\ell >0$ part of $\psi$ takes the form
\begin{eqnarray}
{\psi_{\ell >0}} &=& {\sum _{\ell >0}} {c_\ell} {g_\ell}(b) {h_\ell}(r) {C^\alpha _\ell}(\cos \theta)\; \; \; r<b
\nonumber 
\\
{\psi_{\ell >0}} &=& {\sum _{\ell >0}} {c_\ell} {h_\ell}(b) {g_\ell}(r) {C^\alpha _\ell}(\cos \theta) \; \; \; r>b
\label{psihigher}
\end{eqnarray} 
for some set of constants $c_\ell$.

Now for each $\ell >0$ multiply eqn. (\ref{psi}) by ${\sqrt g} {C^\alpha _\ell}(\cos \theta) $ and integrate over all angular variables to obtain
\begin{equation}
- 4 \pi e {C^\alpha _\ell}(1) \delta (r-b) = {{\cal A}_{n-2}} {Q^\alpha _\ell} \left [ 
{\frac d {dr}} \left ( {r^{n-1}} {\frac {d {A_\ell}} {dr}} \right ) - \ell (\ell + n - 2) {r^{n-3}} {f^{-1}} {A_\ell} \right ]
\label{deltar}
\end{equation}
Ingegrating eqn. (\ref{deltar}) from $b-\epsilon$ to $b+\epsilon$ we obtain
\begin{equation}
- 4 \pi e {C^\alpha _\ell}(1)  = {{\cal A}_{n-2}} {Q^\alpha _\ell} {c_\ell} {b^{n-1}} W({h_\ell},{g_\ell})
\label{cl1}
\end{equation}
Here the Wronskian of two solutions $W({u_1},{u_2})$ is defined to be $W \equiv {u_1}{u_2}' - {u_2}{u_1}'$ and it is to be evaluated at $r=b$.  However, since the differential equation that the solutions satisfy is eqn. (\ref{Aeqn2}) we obtain
\begin{equation}
{\frac {dW} {dr}} = - {\frac {n-1} r} W
\end{equation} 
and therefore there is a constant $k$ for which $W = k {r^{1-n}}$.  But with our chosen normalization for $g_\ell$ and $h_\ell$ we find that at large distances $W = -(2 \ell + n - 2) {r^{1-n}}$ and therefore that the constant $k$ is equal to $- (2 \ell + n - 2)$.  Using this result in eqn. (\ref{cl1}) we obtain
\begin{equation}
- 4 \pi e {C^\alpha _\ell}(1)  = - (2 \ell + n - 2) {{\cal A}_{n-2}} {Q^\alpha _\ell} {c_\ell} 
\label{cl2}
\end{equation}
and therefore
\begin{equation}
{c_\ell} = {\frac {4 \pi e {C^\alpha _\ell}(1)} {(2 \ell + n - 2) {{\cal A}_{n-2}} {Q^\alpha _\ell}}}
\label{cl3}
\end{equation}
(Note that for the case $n=3$ the expression of eqn. (\ref{cl3}) becomes ${c_\ell} = e$).  Using eqn. (\ref{cl3}) in eqn. (\ref{psihigher}) we find that the $\ell >0$ part of $\psi$ is given by
\begin{equation}
{\psi_{\ell >0}} = {\frac {4\pi e} {{\cal A}_{n-2}}} {\sum _{\ell >0}} 
{\frac {{C^\alpha _\ell}(1)} {(2 \ell + n - 2){Q^\alpha _\ell}}} \, {g_\ell}(b) {h_\ell}(r) {C^\alpha _\ell}(\cos \theta)\; \; \; r<b
\end{equation}
\begin{equation}
{\psi_{\ell >0}} = {\frac {4\pi e} {{\cal A}_{n-2}}} {\sum _{\ell >0}} {\frac {{C^\alpha _\ell}(1)} {(2 \ell + n - 2){Q^\alpha _\ell}}} \, {h_\ell}(b) {g_\ell}(r) {C^\alpha _\ell}(\cos \theta) \; \; \; r>b
\label{psihigher2}
\end{equation}

To obtain explicit expressions for $\psi _{\ell >0}$ we need explicit expressions for ${g_\ell}(r)$ and ${h_\ell}(r)$.  However, there is already enough information in eqn. (\ref{psihigher2}) to work out the fate of the higher multipole field as the charge is lowered to the horizon.  Since ${h_\ell}(r)$ vanishes on the horizon, it follows that ${h_\ell}(b)$ goes to zero as the charge is lowered to the horizon.  Therefore in this limit the right hand side of eqn. (\ref{psihigher2}) vanishes.  Thus all higher multipole parts of the field vanish and only the Coulomb field of eqn. (\ref{Coulomb}) remains.

\section{Solutions of the radial equation}
\label{radial}

We now turn to the problem of obtaining explicit expressions for ${g_\ell}(r)$ and ${h_\ell}(r)$.
Since $g_\ell$ behaves like $r^{-(\ell + n - 2)}$ near infinity, we define ${\tilde A}_\ell$ by 
${{\tilde A}_\ell} \equiv {r^{\ell + n - 2}}{A_\ell}$ and find that eqn. (\ref{Aeqn2}) takes the form
\begin{equation}
{\frac {{d^2}{{\tilde A}_\ell}} {d{r^2}}} \; - \; {\frac {2 \ell + n - 3} r} \; {\frac {d{{\tilde A}_\ell}} {dr}} \; + \; {\frac {\ell (\ell + n - 2)} {r^2}} \, (1 - {f^{-1}} ) {{\tilde A}_\ell} = 0 \; \; \; .
\label{Aeqn3}
\end{equation}
Defining the coordinate $\rho \equiv 1-f$ we find that eqn. (\ref{Aeqn3}) takes the form
\begin{equation}
\rho (\rho - 1) {\frac {{d^2}{{\tilde A}_\ell}} {d{\rho ^2}}} \; + \; (\rho - 1) (2s+2) 
{\frac {d {{\tilde A}_\ell}} {d\rho}} \; + \; s(s+1) {{\tilde A}_\ell} = 0
\label{Aeqn4}
\end{equation}
where the quantity $s$ is defined by 
\begin{equation}
s \equiv {\frac \ell {n-2}} \; \; \; .
\label{sdef}
\end{equation}
Note that $r \to \infty$ corresponds to $\rho=0$ and the horizon is at $\rho=1$.  Thus we are interested in solutions to eqn. (\ref{Aeqn3}) on the interval $(0,1)$.  Furthermore, $g_\ell$ is the solution that vanishes at $\rho=0$ and $h_\ell$ is the solution that vanishes at $\rho=1$.  

Eqn. (\ref{Aeqn4}) has the form of the hypergeometric differential equation.  Recall \cite{Grad} that the hypergeometric differential equation for a function $y(x)$ has three parameters 
$({a_1},{a_2},{a_3})$ and takes the form
\begin{equation}
x(x-1) \; {\frac {{d^2}y} {d{x^2}}} \; + \; \left [ ({a_1}+{a_2}+1) x - {a_3}\right ] \; 
{\frac {dy} {dx}} + {a_1}{a_2} y = 0 \; \; \; .
\label{hyper}
\end{equation}
Furthermore, the solution to the hypergeometric equation that is regular at $x=0$ is the hypergeometric function $F({a_1},{a_2},{a_3},x)$.   Comparing eqn. (\ref{Aeqn4}) to eqn. (\ref{hyper}) we find that the values of the parameters are 
\begin{equation}
{a_1} = s, \; \; \; {a_2}=1+s, \; \; \; {a_3}=2+2s
\end{equation}  
It then follows that $g_\ell$ is given by
\begin{equation}
{g_\ell} = {r^{-(\ell + n - 2)}} \, F(s,1+s,2+2s,\rho) \; \; \; .
\label{gsoln}
\end{equation}
Since $F({a_1},{a_2},{a_3},0)=1$, it follows that eqn. (\ref{gsoln}) has the normalization for $g_\ell$ that we chose in the previous section.

We could attempt to find $h_\ell$ by using a linear combination of the singular solution and the non-singular solution of eqn. (\ref{Aeqn4}).  However, it turns out to be both easier and more straightforward to use $f$ as a variable instead of $\rho$ and to build in the property that $h_\ell$ needs to vanish at the horizon: from eqn. (\ref{Aeqn4}) we obtain
\begin{equation}
f(f-1) \; {\frac {d^2} {d{f^2}}} \; ( {f^{-1}} {{\tilde A}_\ell} ) \; + \; \left [ (2s+4) f - 2 \right ] \; {\frac d {df}} \; ( {f^{-1}} {{\tilde A}_\ell} ) \; + \; (s+1)(s+2) {f^{-1}} {{\tilde A}_\ell} = 0 \; \; \; .
\label{Aeqn5}
\end{equation}
Equation (\ref{Aeqn5}) is also the hypergeometric equation, but now with the parameters
\begin{equation}
{a_1} = 1+s, \; \; \; {a_2}=2+s, \; \; \; {a_3}=2
\end{equation} 
Taking the nonsingular solution, we then find that $h_\ell$ is given by 
\begin{equation}
{h_\ell} = {k_\ell} {r^{-(\ell + n - 2)}} \, f \, F(1+s,2+s,2,f) \; \; \; .
\label{hsoln}
\end{equation}
Here $k_\ell$ is a normalization constant to be chosen to satisfy the normalizaton condition chosen in the previous section.

\section{Conclusion}
\label{conclusion}
We have found that all the higher multipole moments vanish as the charge is lowered to the horizon.  What then went wrong in the analysis of \cite{fox} to yield the opposite conclusion?  Simply put, the treatment of \cite{fox} chooses solutions of Maxwell's equations that are singular on the horizon, with that choice being obscured by the coordinate systems used.  The method of \cite{fox} uses the $t$ coordinate throughout, and uses the $\rho$ coordinate to analyze all solutions of the radial equation, which makes for a very complicated analysis at the horizon.  Using the Eddington coordinate $v$, one can immediately see from eqn. (\ref{potential}) that the higher multipole part of $\psi$ must vanish at the horizon.  But in any coordinate system, one can calculate invariant quantities and demand that they be nonsingular.  From eqns. (\ref{vrmetric2}) and (\ref{potential}) it follows that the electromagnetic invariant ${F^{ab}}{F_{ab}}$ is given by
\begin{equation}
{F^{ab}}{F_{ab}} = -2 \left [  {{({\partial _r}\psi )}^2} + {f^{-1}} {r^{-4}} {{({\partial _\theta}\psi )}^2}  \right ] \; \; \; .
\end{equation} 
Therefore, from an examination of this invariant one can conclude that the non-monopole part of $\psi$ must vanish on the horizon.  The treatment of \cite{fox} fails to impose this condition and is therefore not treating the correct electromagnetic field.

In contrast, we impose smoothness on the horizon and find that everything proceeds as a straightforward generalization of \cite{wald} with the same conclusion: all higher multipoles vanish as the charge approaches the horizon.  There may be many cases in which higher dimensional black holes lead to exotic, unexpected, behavior, but this is not one of them.

\section*{Acknowledgments}

This work was supported by NSF Grant PHY-1806219.

\end{document}